\documentclass[letter,traditabstract]{aa} 
\usepackage{grffile}
\usepackage[varg]{txfonts}
\usepackage{subfigure}
\usepackage{amssymb}
\usepackage{cancel}
\usepackage{bm}
\usepackage{hyperref}
\usepackage{color}
\usepackage{footnote}
\usepackage{eqnarray}
\usepackage{verbatim}
\begin{document}

   \title{High gas-to-dust size ratio indicating efficient radial drift in the mm-faint CX Tau disk}
   
\titlerunning{High gas/dust size ratio in the CX Tau disk}

   \author{S. Facchini\inst{1,2}   
           \and
           E. F. van Dishoeck\inst{2,3}
           \and
           C. F. Manara\inst{1}
           \and
           M. Tazzari\inst{4}
           \and
           L. Maud\inst{1,3}
          \and
           P. Cazzoletti\inst{2}
           \and
           G. Rosotti\inst{3}
           \and
           N. van der Marel\inst{5}
           \and
           P. Pinilla\inst{6}
           \and
           C. J. Clarke\inst{4}
          }

   \institute{European Southern Observatory, Karl-Schwarzschild-Str. 2, 85748 Garching, Germany,\\
            \email{stefano.facchini@eso.org}
             \and
             Max-Planck-Institut f\"ur Extraterrestrische Physik, Giessenbachstrasse 1, 85748 Garching, Germany
             \and
             Leiden Observatory, Leiden University, Niels Bohrweg 2, NL-2333 CA Leiden, The Netherlands
             \and
             Institute of Astronomy, Madingley Road, Cambridge, CB3 OHA, UK
             \and
             National Research Council of Canada Herzberg Astronomy and Astrophysics Programs, 5071 West Saanich Road, Victoria, BC, V9E2E7, Canada
             \and
             Department of Astronomy/Steward Observatory, The University of Arizona, 933 North Cherry Avenue, Tucson, AZ 85721, USA
             }

   \date{Received; accepted}

 
  \abstract{The large majority of protoplanetary disks have very compact  continuum emission ($\lesssim15\,$AU) at millimeter wavelengths. However, high angular resolution observations that resolve these small disks are still lacking, due to their intrinsically fainter emission compared with large bright disks. In this letter we present $1.3\,$mm ALMA data of the faint disk ($\sim10\,$mJy)  orbiting the TTauri star CX Tau at a resolution of $\sim40\,$mas, $\sim5\,$AU in diameter. The millimeter dust disk is compact, with a 68\% enclosing flux radius of 14\,AU, and the intensity profile exhibits a sharp drop between 10 and 20\,AU, and a shallow tail between 20 and 40\,AU. No clear signatures of substructure in the dust continuum are observed, down to the same sensitivity level of the DSHARP large program. However, the angular resolution does not allow us to detect substructures on the  scale of the disk aspect ratio in the inner regions. The radial intensity profile closely resembles  the inner regions of more extended disks imaged at the same resolution in DSHARP, but with no rings present in the outer disk. No inner cavity is detected, even though the disk has been classified as a transition disk from the spectral energy distribution in the near-infrared. The emission of $^{12}$CO is much more extended, with a 68\% enclosing flux radius of 75\,AU. The large difference of the millimeter dust and gas extents ($>5$) strongly points to  radial drift, and closely matches  the predictions of theoretical models.}

   \maketitle
%
\section{Introduction}

In the last few years, the Atacama Large Millimeter/submillimeter Array (ALMA) has conducted surveys of several nearby star forming regions, targeting hundreds of protoplanetary disks previously selected from spectral classification in the infrared (IR)  \citep[e.g.,][]{2016ApJ...828...46A,2017AJ....153..240A,2016ApJ...831..125P,2016ApJ...827..142B,2019MNRAS.482..698C}. These shallow surveys allowed the derivation of statistical properties of protoplanetary disks, which are crucial when constraining the evolution of these systems and probing the planet formation potential (past or present) of young stellar objects.

One of the many interesting results of these surveys is that many disks show rather compact continuum emission at (sub-)millimeter (hereafter mm) wavelengths. In particular, star forming regions like Lupus and Ophiucus present many objects ($\gtrsim40\%$ of the detected sources) that are still spatially unresolved at $\sim0.2\arcsec$ resolution ($\sim30\,$AU in diameter). These compact objects are also the faintest in the mm luminosity distribution, as expected from the luminosity--size relation derived from the same surveys \citep{2017ApJ...845...44T,2017A&A...606A..88T,2018ApJ...865..157A}. This result is supported by earlier photometric studies of spectral energy distributions (SEDs), suggesting that many disks can be as small as $1.5\,$AU or less \citep[e.g.,][]{2017ApJ...841..116H,2018MNRAS.477..325B}.

\begin{figure*}
\begin{center}
\includegraphics[width=0.75\textwidth]{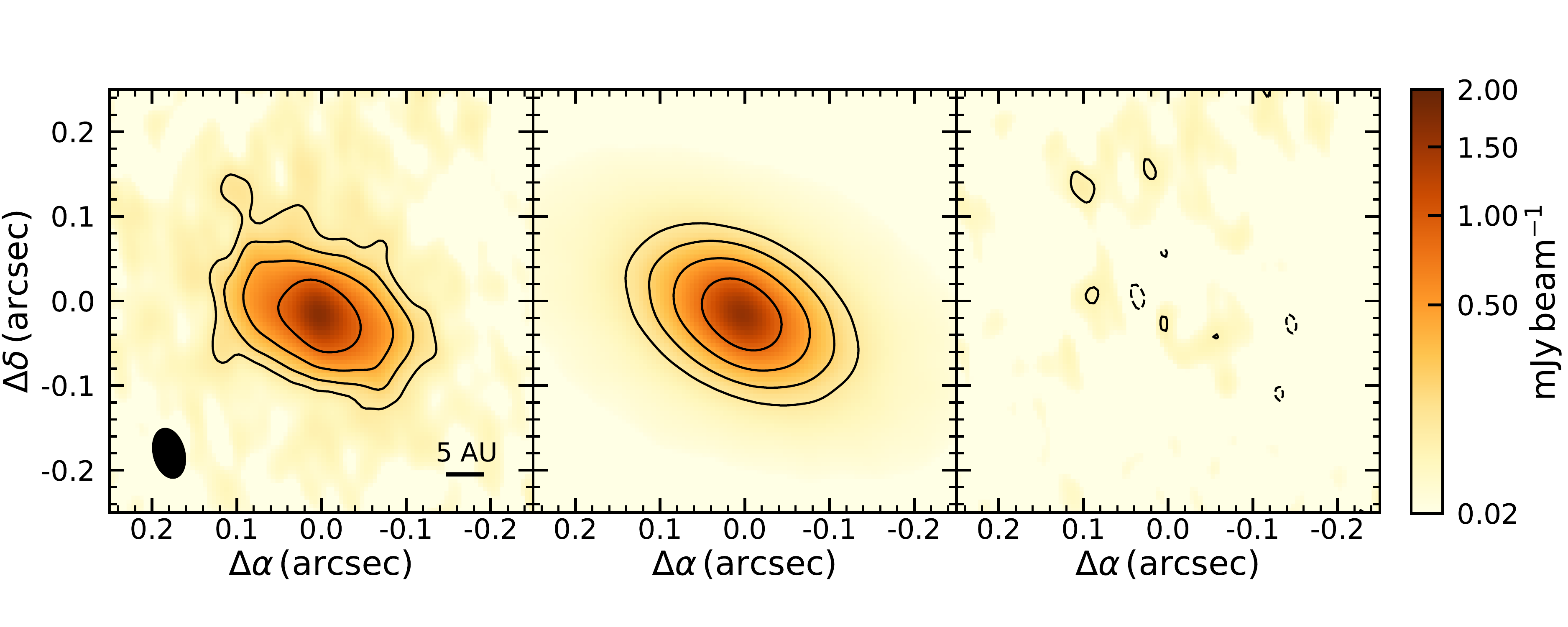}
\end{center}
\caption{From left to right: Data, best fit model, and residuals of the continuum measurement set. The model image and residuals were both  created with best fit model, and residual visibilities with the same cleaning parameters as for the data image. Contour levels are at $[5,10,20,40]\sigma$ in the first two panels, and $[-3,3]\sigma$ in the map of the residuals. The synthesized beam is shown in the bottom left corner of the left panel.}
\label{fig:map}
\end{figure*}

The high occurrence of compact disks is expected from radial drift models, where dust pebbles drift towards (local) pressure maxima present in the gas structure \citep[e.g.,][]{1977MNRAS.180...57W}. In case of smooth pressure profiles, the dust is expected to accumulate in the inner regions of the disks. These theoretical predictions have been challenged by objects showing very extended continuum emission at mm wavelengths. However, recent high angular resolution observations of bright disks have shown that most (if not all) of these objects host radial substructures \citep[e.g.,][]{2015ApJ...808L...3A,2016ApJ...820L..40A,2018ApJ...869L..41A,2018ApJ...866L...6C,2018ApJ...869...17L,2018A&A...616A..88V}, suggesting that radial drift accumulates dust pebbles at local pressure maxima at large radii \citep[e.g.,][]{2012A&A...538A.114P,2018ApJ...869L..46D}. Gaps in gas surface density and pressure gradients from the gas kinematics have shown in some cases that disks hosting rings and gaps in the dust structure also present a non-monotonic pressure profile \citep[e.g.,][]{2017A&A...600A..72F,2018ApJ...860L..12T}. However, high angular resolution observations have mostly focused on bright (and thus large) protoplanetary disks, whereas very little is known about radially resolved properties of compact objects. In particular, it is not clear whether the same level of substructures is present on smaller scales in compact disks. 

The theoretical expectations of radial drift are that the dust surface density should be less extended than the gas \citep{birnstiel_14} and that the disk's average dust-to-gas ratio should be lower than the canonical 1\%, with the inner regions locally showing a higher value. Both effects predict that the mm continuum emission should be much more compact than the emission in gas tracers, in particular CO \citep[e.g.,][]{2017A&A...605A..16F,2019arXiv190306190T}. Initial surveys of the gas and dust radii indicate that the majority of objects that are bright in the mm continuum exhibit moderate difference in the radial extent of CO and continuum, within a factor of $1.5-3$ \citep[e.g.,][]{2017ApJ...851...85B,2018ApJ...859...21A}. For a few cases some amount of radial drift may be required to explain the observed difference \citep[e.g.,][]{2009A&A...501..269P,2012ApJ...744..162A}; instead,  the majority of these observations can be explained without invoking radial drift, with the difference in the emission extent caused by the opacity in CO being higher than in the dust \citep[e.g.,][]{1998A&A...338L..63D,1998A&A...339..467G,2017A&A...605A..16F}. It is still unclear whether the sources showing compact mm continuum emission are scaled-down versions of the large disks, with the gas component also being more compact than for large disks, or whether the gas in these objects extends to large radii, with a difference in the gas and dust extents that is $\gg3$. Observational evidence of such a difference would be a clear indication of radial drift \citep{2017A&A...605A..16F,2019arXiv190306190T}. 

CX Tau is a good target that can be used to characterize the gas and dust properties (in particular the radial extent) of a compact continuum protoplanetary disk at high angular resolution due to its relatively low mm flux. CX Tau is an M2.5 star, located at $127.9\pm0.7\,$pc from the Sun \citep{2016A&A...595A...1G,2018A&A...616A...1G}, with an effective temperature $T_{\rm eff}=3483\pm48\,$K and a bolometric luminosity $\log{L_*/L_\odot}=-0.66\pm0.20$ \citep{2014ApJ...786...97H}. The dynamical estimate for the mass of the star is $0.31\,M_\odot$ \citep{2017ApJ...844..158S}, in excellent agreement with the mass from stellar evolution models. The star has an accretion rate of $\dot{M}=7.1\times10^{-10}\,M_\odot\,$yr$^{-1}$ \citep{1998ApJ...495..385H}. The luminosity and accretion rate values have been updated using the new Gaia distance. A disk has been  marginally resolved in continuum with the IRAM Plateau de Bure interferometer at 1.3\,mm \citep{2014A&A...564A..95P}, with an integrated flux density of $9.6\pm0.2\,$mJy, which characterizes CX Tau as an average object when compared to the recent ALMA surveys. The disk is also clearly detected and resolved in the $^{12}$CO $J$=2-1 line \citep{2017ApJ...844..158S}, at $\sim0.25\arcsec$ resolution with ALMA.  

In this letter, we present significantly higher angular resolution ALMA data of CX Tau at 1.3\,mm that provide new insights into the properties of its gas and dust components. Sensitivity and angular resolution are as in the DSHARP large program \citep{2018ApJ...869L..41A}. Section \ref{sec:observations} describes the observations, Section \ref{sec:results} shows the analysis and results, and Section \ref{sec:discussion} discusses these results within the broader context of disk evolution.

\section{Observations and data reduction}
\label{sec:observations}

CX Tau was observed in two configurations (C40-5 and C40-8) on 2016 November 5 and 2017 September 25, respectively (ALMA Program \#2016.1.00715.S, PI Facchini). The compact configuration spanned baselines between 18 and 1124\,m, while the extended configuration ranged between 41 and 14851\,m. Both observations were performed with 40 antennas, with on-source integration times of 20 and 54 minutes, and water vapor levels of 1.72\, mm and 0.35\,mm, respectively. In the C40-5 configuration, J0510+1800 was used as both bandpass and flux calibrator. In the C40-8 configuration, the same calibrator was used for bandpass, whereas J0238+1636 was used as flux calibrator. In both observations J0403+2600 was chosen as phase calibrator. Flux calibrators were quantified within a week for both observations. The spectral setup had seven spectral windows, with the following combination of central frequency and bandwidth: (217.238\,GHz, 234.\,MHz); (217.824\,GHz, 234\,MHz); (218.222\,GHz, 468\,MHz); (219.560\,GHz, 58\,MHz); (220.398\,GHz, 58\,MHz); (230.538\,GHz, 117\,MHz); (233.000\,GHz, 1875\,MHz). The spectral setup was chosen to have a full spectral window dedicated to continuum
, and the other spectral windows targeting different molecular lines. This work presents only the observations of the $^{12}$CO line, with a resolution of 61 kHz (0.079\,km\,s$^{-1}$). All spectral windows were used for continuum, excluding all channels within $\pm 60\,$km\,s$^{-1}$ from the targeted lines. The total bandwidth for the continuum results in $2661\,$MHz.

\begin{figure}
\center
\includegraphics[width=\columnwidth]{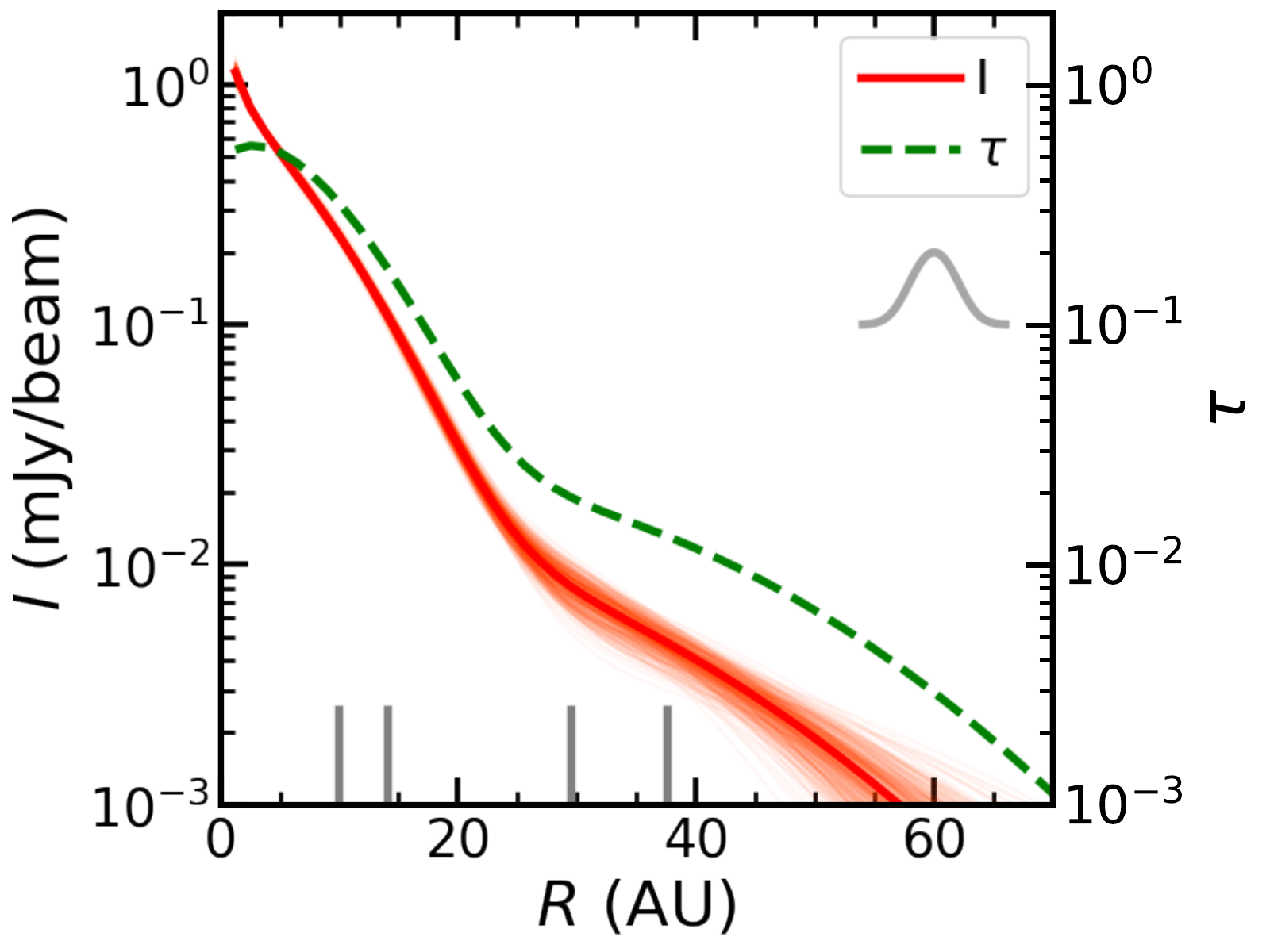}
\caption{Intensity and optical depth profiles of the best fit model. The solid red line shows the best fit radial intensity profile of the continuum;  400 random realizations were chosen from the posterior distribution of the fit to highlight the uncertainties. The green dashed line indicates the optical depth profile of the continuum assuming the temperature profile from equation~\ref{eq:temp}. The gray vertical lines show the radii enclosing 50\%, 68\%, 90\%, and 95\% of the total flux. The Gaussian in the top right corner has the same FWHM of the minor axis of the beam.}
\label{fig:profile}
\end{figure}

The data were calibrated using the {\tt CASA} package, version 4.7 \citep{2007ASPC..376..127M}. Self-calibration was performed on the datasets for both phase and amplitude, with solution intervals equal to the scan length and 120\,s for the compact configuration data, and to the scan length and 360\,s for the extended configuration data. This procedure led to a modest improvement in the signal-to-noise ratio (S/N) in the continuum and in the $^{12}$CO line. The visibilities were then merged using the {\tt concat} task in {\tt CASA}. Images in the sky plane were produced using the {\tt clean} task. Different weighting schemes were tested to produce the images. The best compromise between angular resolution and S/N with the continuum data is with a Briggs robust weighting  of 0.5. The resulting synthesized beam is $55\times32\,$mas, with a position angle (P.A.) of $13.6^\circ$. The rms noise level is $\sim20.3\,\mu$Jy\,beam$^{-1}$. For the $^{12}$CO line, the data were re-binned into $0.1\,$km\,s$^{-1}$ channels. Due to the lower S/N, the line was imaged with a Briggs robust weighting  of 2.0, and an additional $0.1\arcsec$ taper. The restoring beam is $0.17\arcsec\times0.13\arcsec$, with a P.A. of $23.5^\circ$. The rms noise level is $\sim4.4\,$mJy\,beam$^{-1}$ in one $0.1\,$km\,s$^{-1}$ channel.

\section{Results}
\label{sec:results}

\subsection{Radial extent of continuum}
\label{sec:results_cont}

The continuum disk is clearly small, and is well resolved at this resolution (Fig.~\ref{fig:map}). To characterize the intensity profile of the mm emission, the disk is modeled in the $uv$-plane assuming azimuthal symmetry. To do so, we use the GALARIO code \citep[][see details in the Appendix]{2018MNRAS.476.4527T}. A best model is obtained by performing the fit using a Markov chain Monte Carlo (MCMC) sampler.
\begin{figure}
\center
\includegraphics[width=.95\columnwidth]{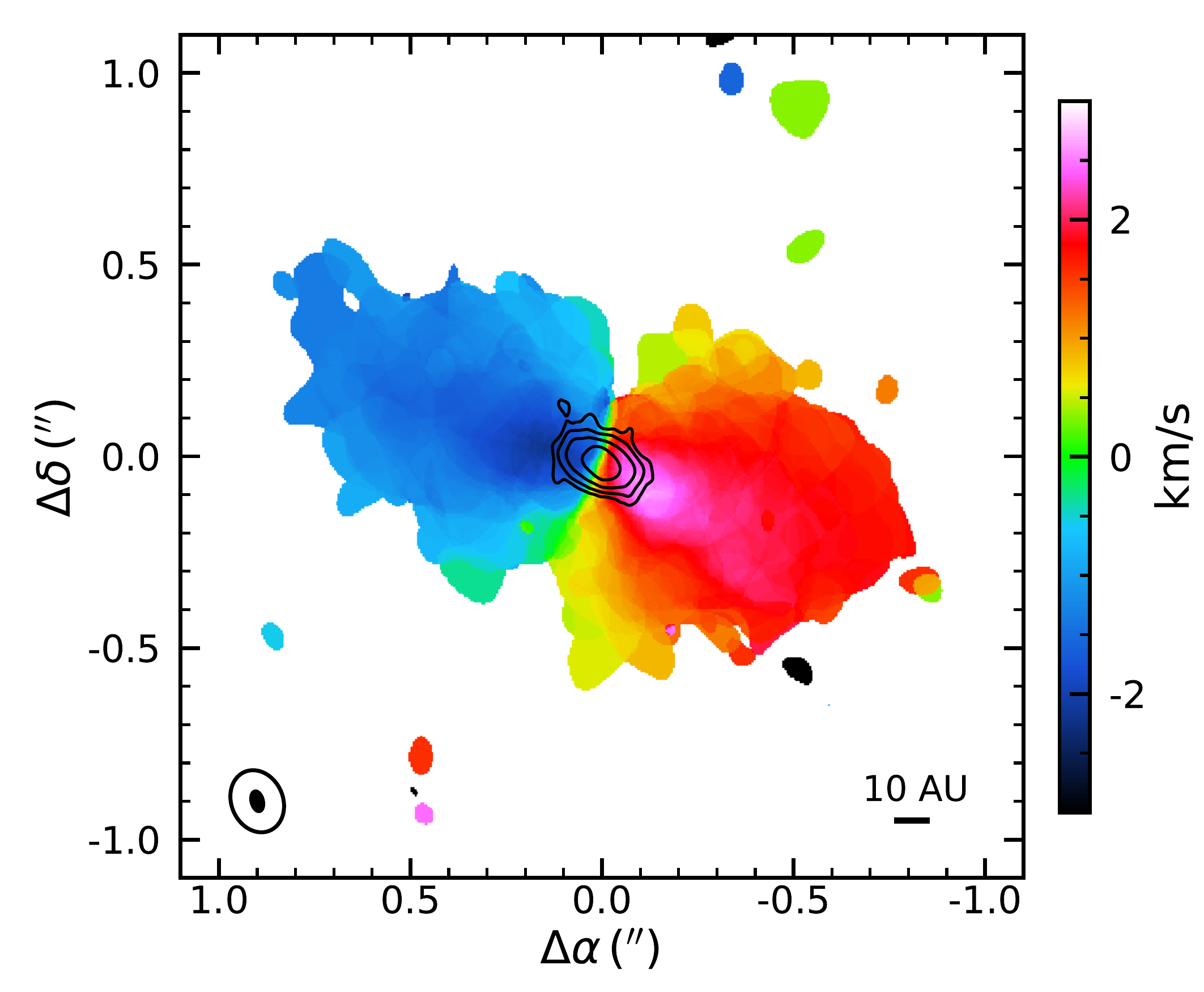}
\caption{Intensity weighted velocity (moment 1) map of the $^{12}$CO $J$=2-1 line, where the flux density in the channels has been clipped at $4\sigma$. Contours show the continuum emission, with contour levels at $[5,10,20,40]\sigma$. The beams of continuum (full ellipse) and CO line (empty ellipse) are shown in the bottom left corner.}
\label{fig:mom1}
\end{figure}

Different functional forms for the radial profile of the intensity were attempted
. By analysing the azimuthally averaged profile obtained from the image, we decided to use the  functional form

\begin{equation}
I(R) = I_0 \left(\frac{R}{\sigma_0}\right)^{-\gamma}e^{-R^2/2\sigma_0^2}+I_1e^{-R^2/2\sigma_1^2},
\label{eq:intensity}
\end{equation}
where $I_0$ and $I_1$ are normalization terms, $\sigma_0$ and $\sigma_1$ characterize two Gaussians centered in the disk origin, and $R$ is the cylindrical radial coordinate. Together with the five parameters describing the radial profile, the fit is performed over the inclination of the disk, its position angle, and the right ascension (RA) and declination (Dec) offsets between the disk center and the observation phase center. The re-centered and de-projected visibilities and best fit model (i.e., the maximum likelihood model) are presented in the Appendix, whereas the derived intensity profile is shown in Fig.~\ref{fig:profile}. The best fit parameters all fall within the central $68\%$ interval of the marginalized posterior distributions
. The derived integrated flux density is $9.75\pm0.12\,$mJy, in agreement with \citet{2014A&A...564A..95P}.

The extent of the continuum emission is computed from the inferred posteriors. The radius enclosing 68\% of the continuum flux density is $14.0\pm0.3\,$AU, where the error is computed as the statistical uncertainty. Table~\ref{tab:radii} in the Appendix lists the radii extracted at typical flux density levels used in the literature. CX Tau falls within the mm size--luminosity  relation by \citet{2018ApJ...865..157A}.

\subsection{Radial extent of CO}

The $^{12}$CO $J$=2-1 line is clearly detected. The intensity weighted velocity (moment 1) map is presented in Fig.~\ref{fig:mom1}. The integrated flux of the line is $1937\pm54\,$mJy\,km\,s$^{-1}$, as computed in the image plane within an elliptical area with semimajor axis $1.53\arcsec$. Inclination, P.A., and center of the ellipse are taken from the continuum best fit parameters. The aperture radius is determined via a curve-of-growth method, with successively larger elliptical apertures until the incremental flux is less than the standard deviation of 20 apertures taken away from the source. With the same method, the azimuthally averaged intensity profile is computed, as in \citet{2018ApJ...859...21A}. The radius enclosing $68\%$ of the $^{12}$CO $J$=2-1 flux is $75\pm$5\,AU. Radii enclosing different percentages of the total flux are listed in Table~\ref{tab:radii}. 

The difference in radial extent between continuum and line emission is remarkable. At the $68\%$ enclosing flux radius the ratio of  $R_{\rm gas}$ to $R_{\rm dust}\approx5.4$, while at the $90\%$ enclosing flux radius the ratio is $\approx3.9$. \citet{2018ApJ...859...21A} performed a similar study on a larger sample of disks in the Lupus region using the 90\% enclosing flux radius; they found that the typical ratio is $1.96$, with a standard error on the mean of $0.04$. In all cases $R_{\rm gas}<3R_{\rm dust}$. However, their study is biased to large and bright objects due to the short integration ($\sim1\,$min at $0.25\arcsec$ resolution). There are a few objects where the gas extent is notably larger than the mm continuum extent \citep[e.g.,][]{2013ApJ...775..136R}, but they do not seem to be typical at the high flux end of the disks distribution.

\subsection{Continuum intensity profile}
\label{sec:res_3}

The continuum intensity profile does not show any significant level of substructure at the angular resolution of the observations (see Fig.~\ref{fig:map}). The visibilities do show some hint of substructures at the $2\sigma$ level that the model is not able to reproduce. Some attempts have been made to refine the model by adding rings and gaps, but no better constraint was obtained. The synthesized image of the residual visibilities are shown in Fig.~\ref{fig:map}, where the residuals are clearly low in flux ($3.3\sigma$ at maximum), and do not show a coherent pattern.

The scarcity of rings and gaps (or other substructures) in CX Tau at this resolution can be due to three factors: 1) substructures are present, but not resolved; 2) surface density and opacity are smooth; and 3) the optical depth in the very inner regions is so high that variations in opacity or surface density are weakly imprinted onto the intensity profile. In order to derive a simple estimate of the optical depth, we assume a simple mid-plane radial temperature profile, as in \citet{2018ApJ...869L..42H}

\begin{equation}
T_{\rm mid}(R)= \left( \frac{\varphi L_{*}}{8\pi R^2\sigma_{SB}} \right)^{0.25},
\label{eq:temp}
\end{equation}
where $\sigma_{\rm SB}$ in the Stefan-Boltzmann constant, and $\varphi$ is the flaring angle, which we set equal to $0.02$ as in \citet{2018ApJ...869L..42H}. The optical depth $\tau_\nu$ was derived using the relation

\begin{equation}
I_{\nu}(R) = B_{\nu}(T_{\rm mid}(R)) (1-\exp{(-\tau_\nu(R)}),
\end{equation}
where $B_{\nu}$ is the blackbody intensity. For the intensity profile we used the best fit model. The derived optical depth is shown in Fig.~\ref{fig:profile}. Only the very inner region ($R<10\,$AU) appears marginally optically thick (the peak of $\tau$ being $\sim0.5$), with the optical depth decreasing rapidly at larger radii. Whether the optical depth profile is affected by unresolved narrow gaps and rings is difficult to judge. A high angular resolution spectral index map would definitely help interpreting the results. The unresolved spectral index between 1 and 3\, mm is $2.3$ \citep{2010A&A...512A..15R}, suggestive of high optical depth or significant grain growth in the regions dominating the flux, i.e., the inner 10\,AU \citep[e.g.,][]{2014prpl.conf..339T}. Nevertheless, the angular resolution limits the possibility of distinguishing between a smooth surface density profile with intermediate optical depth in the inner 10\,AU and unresolved optically thick substructures. \citet{2018ApJ...869L..42H} noted that when rings and gaps occur, the associated radial width is of the order of the pressure scale height of the disk (with some exceptions). Even an unrealistically hot disk, with an aspect ratio $H/R\sim0.1$, would show rings with a typical width of $\lesssim0.5\,$AU in the central core of CX Tau, which cannot be observed with the present antenna configuration.

\begin{figure}
\center
\includegraphics[width=\columnwidth]{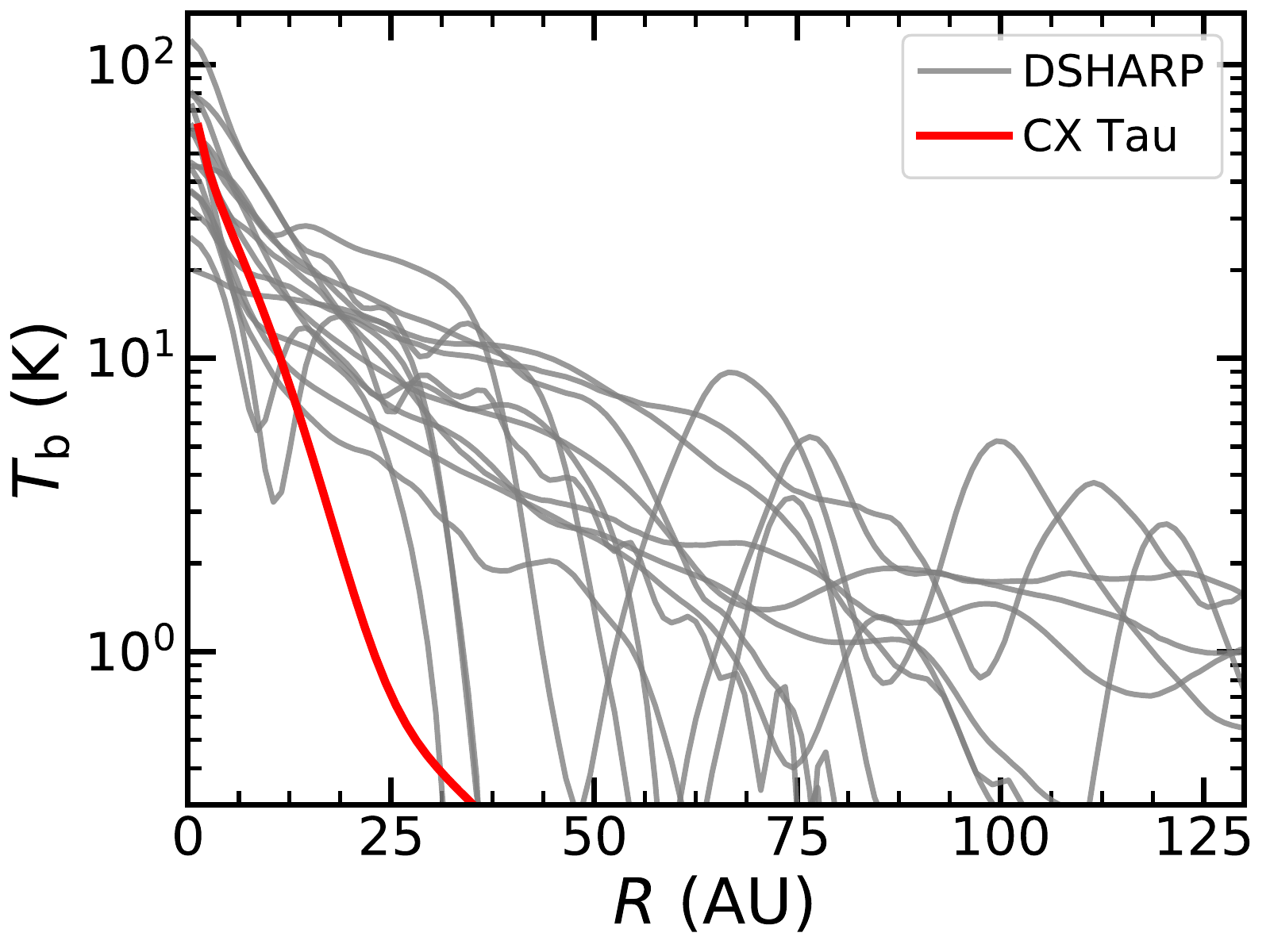}
\caption{Continuum brightness temperature profiles of the best fit of CX Tau (red line) and all DSHARP disks that are not in binary systems and do not show an inner cavity \citep[data from][]{2018ApJ...869L..41A,2018ApJ...869L..42H}. The brightness temperature is computed from the intensity map in the Rayleigh-Jeans approximation using source specific distances from the Gaia catalogue.}
\label{fig:DSHARP}
\end{figure}

It is informative to compare the continuum intensity profile of the CX Tau disk with the profiles of the DSHARP large program \citep{2018ApJ...869L..41A,2018ApJ...869L..42H}. Figure~\ref{fig:DSHARP} compares CX Tau with all the disks from DSHARP that are not in binary systems and that now show an inner cavity. The angular resolution and sensitivity of the two programs match closely. The intensity profile of CX Tau closely resembles  the inner regions of the DSHARP disks, indicating that faint objects are not scaled-down versions of large disks in their continuum emission. They instead exhibit similar intensity profiles, but no outer rings are observed.

Interestingly, CX Tau has been classified as a transition disk from the infrared excess in the SED \citep{2007MNRAS.378..369N}, whereas common infrared color criteria do not include CX Tau in the transition disk definition \citep{2011ApJS..195....3F}. We performed additional analysis on the visibility modeling to verify whether CX Tau hosts a small inner cavity in the dust content. To do so, the additional parameter $R_{\rm trunc}$ was added to the intensity profile of equation~\ref{eq:intensity}, where the intensity was set to 0 for $R<R_{\rm trunc}$. A new fit was performed following the same procedure as before, with 100 walkers in this case. The best fit parameters and marginalized distributions are shown in the Appendix. The data do not show signatures of a cavity, with an upper limit on $R_{\rm trunc}$ of $0.54\,$AU computed as the radius enclosing $95\%$ of the MC realizations. Interestingly, \citet{2012A&A...544A..79B} showed that disks having a near-IR deficit do not need to have a cavity in the mm if the turbulence is so low that the maximum grain size of pebbles throughout the whole disk is set by radial drift rather than fragmentation \citep[an opacity effect, as initially suggested by][]{1989AJ.....97.1451S}.

\section{Discussion and conclusions}
\label{sec:discussion}

\begin{figure}
\center
\includegraphics[width=0.9\columnwidth]{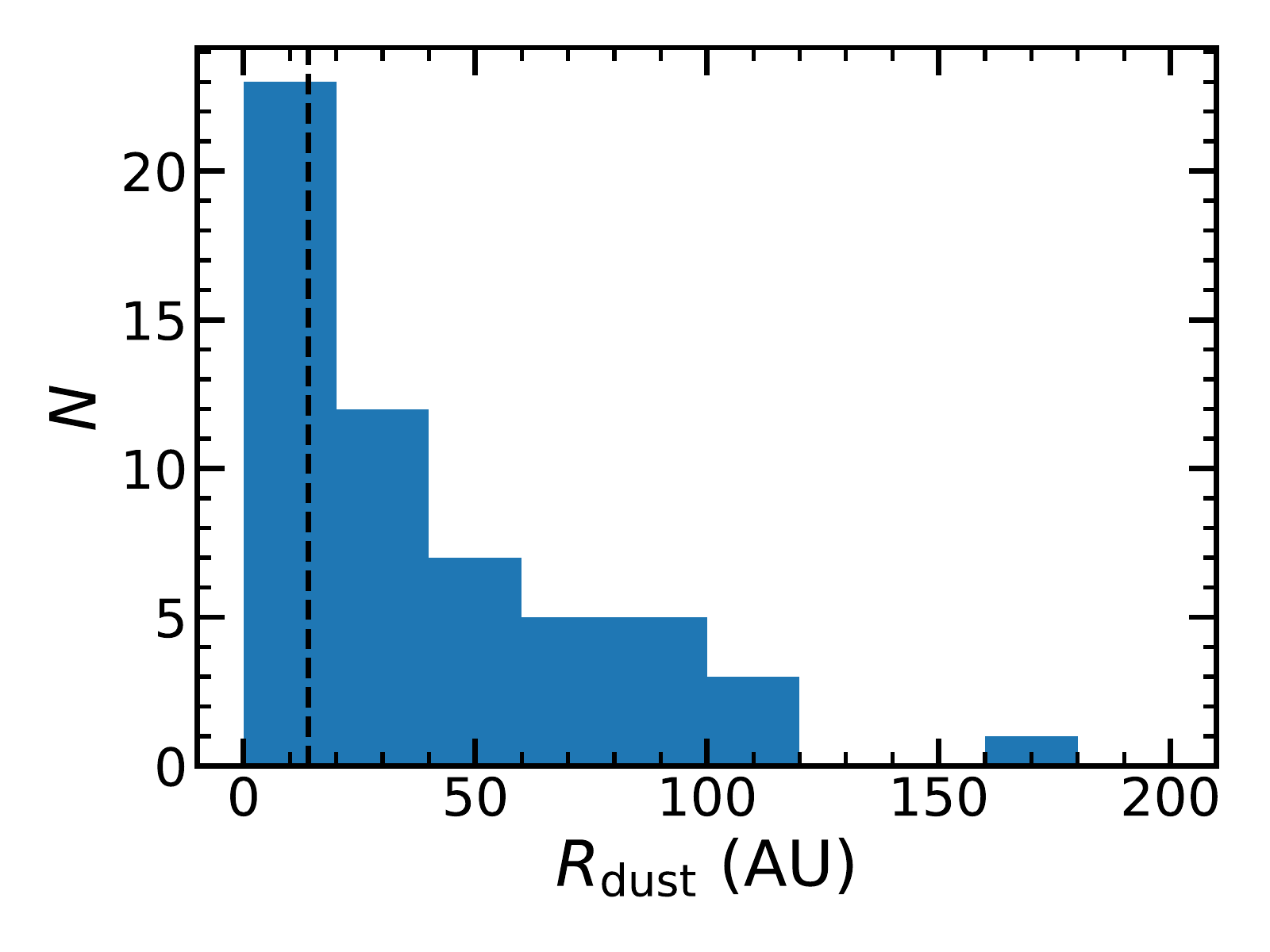}
\caption{Histogram of dust radii (defined as 68\% enclosing radii) in Lupus \citep[data from][]{2016ApJ...828...46A,2018ApJ...865..157A}. The dashed black line indicates CX Tau. The first bin includes unresolved sources.}
\label{fig:histo}
\end{figure}

This letter presents the first high angular resolution ALMA observations of a mm-faint disk in a single (i.e., not binary) system. 
The source is rather compact, and falls within the smallest resolved sources in the Lupus survey \citep[see][and Fig.~\ref{fig:histo}]{2017A&A...606A..88T,2018ApJ...865..157A}, while being in the top $55$\% of detected disk fluxes \citep[e.g., Fig. 4 in][]{2019arXiv190402409C}. This indicates that the majority of Class II disks are even more compact in continuum emission than CX Tau if the size--luminosity relation holds at the low flux end.


The large difference in extent between mm continuum and molecular emission strongly argues for radial drift. \citet{2019arXiv190306190T} show that a ratio of $R_{\rm gas}/R_{\rm dust}\gtrsim4$ cannot be reproduced by invoking optical depth effects only, but does require radial drift. This can be intended either as a smaller radial extent of the dust compared to the gas \citep{birnstiel_14}, or a depletion in the total dust-to-gas ratio which makes grains smaller due to less efficient coagulation \citep{2012A&A...539A.148B} and reduces their sub-mm opacity \citep{2019MNRAS.tmp.1141R}. Realistically, the two effects occur simultaneously. The sharp cliff in mm emission between 10 and 20\,AU may point to the first scenario, as expected if the gas pressure gradient is steep. On the other hand, the sharp dependence of dust opacity with dust size also produces a similar effect \citep{2019MNRAS.tmp.1141R} in a disk with a maximum grain size decreasing steeply with radius. Interestingly, the models by \citet{2019MNRAS.tmp.1141R} also predict the faint shallow drop-off in intensity seen outside 20\,AU, with small grains still contributing to the mm continuum emission in the outer regions. Disentangling between the two effects requires knowing the gas surface density, which cannot be retrieved with the present data.

The observations of CX Tau suggest that objects showing compact (faint) emission at mm wavelengths can still possess fairly extended gaseous disks, where the difference in size between the gas and dust components is due to efficient radial drift. The physical mechanism hindering efficient drift in disks that are bright and extended in mm continuum emission by generating radial perturbations in the gas radial pressure profile may not have occurred in the fainter objects. If dust is trapped by means of quite massive planets in bright objects \citep[e.g.,][]{2018ApJ...864L..26B,2018ApJ...869L..47Z}, the large number of compact disks would suggest that their occurrence at large orbital radii is fairly rare. However, this scenario assumes that rings in the outer disks are long-lived, and directly comparing disks with detected substructures with the exoplanet population should be done with caution \citep[e.g.,][]{2019MNRAS.486..453L}. Objects like CX Tau may instead represent more evolved disks that have lost their outer rings;  the low accretion rate onto CX Tau suggests that this could be a realistic option. Moreover, CX Tau presents a flux in CO (and thus CO radius since CO is mostly optically thick) that is well below the median of the CO fluxes from the DSHARP data. The absence of rings in the outer disk could be simply due to different initial conditions, with large gaseous disks leading to large disks in continuum. Surveys of gas (CO) disk radii are needed to assess how the gas properties of disks relate to the level of substructure in continuum. The difference in radial extent of gas and mm dust in CX Tau suggests that gas outer radii are to be determined independently from mm dust outer radii, in particular to constrain disk evolution \citep[][]{2018ApJ...864..168N}. More high resolution observations of compact disks are needed to confirm whether any substructures in the mm emission can be detected, and whether the remarkable gas-to-dust size ratio is a common feature of the bulk of the disk population.


\bibliographystyle{aa}
\bibliography{references}

\begin{thebibliography}{55}
\expandafter\ifx\csname natexlab\endcsname\relax\def\natexlab#1{#1}\fi

\bibitem[{{ALMA Partnership} {et~al.}(2015){ALMA Partnership}, {Brogan},
  {P{\'e}rez}, {Hunter}, {Dent}, {Hales}, {Hills}, {Corder}, {Fomalont},
  {Vlahakis}, {Asaki}, {Barkats}, {Hirota}, {Hodge}, {Impellizzeri}, {Kneissl},
  {Liuzzo}, {Lucas}, {Marcelino}, {Matsushita}, {Nakanishi}, {Phillips},
  {Richards}, {Toledo}, {Aladro}, {Broguiere}, {Cortes}, {Cortes}, {Espada},
  {Galarza}, {Garcia-Appadoo}, {Guzman-Ramirez}, {Humphreys}, {Jung}, {Kameno},
  {Laing}, {Leon}, {Marconi}, {Mignano}, {Nikolic}, {Nyman}, {Radiszcz},
  {Remijan}, {Rod{\'o}n}, {Sawada}, {Takahashi}, {Tilanus}, {Vila Vilaro},
  {Watson}, {Wiklind}, {Akiyama}, {Chapillon}, {de Gregorio-Monsalvo}, {Di
  Francesco}, {Gueth}, {Kawamura}, {Lee}, {Nguyen Luong}, {Mangum}, {Pietu},
  {Sanhueza}, {Saigo}, {Takakuwa}, {Ubach}, {van Kempen}, {Wootten},
  {Castro-Carrizo}, {Francke}, {Gallardo}, {Garcia}, {Gonzalez}, {Hill},
  {Kaminski}, {Kurono}, {Liu}, {Lopez}, {Morales}, {Plarre}, {Schieven},
  {Testi}, {Videla}, {Villard}, {Andreani}, {Hibbard}, \&
  {Tatematsu}}]{2015ApJ...808L...3A}
{ALMA Partnership}, {Brogan}, C.~L., {P{\'e}rez}, L.~M., {et~al.} 2015, \apjl,
  808, L3

\bibitem[{{Andrews} {et~al.}(2018{\natexlab{a}}){Andrews}, {Huang},
  {P{\'e}rez}, {Isella}, {Dullemond}, {Kurtovic}, {Guzm{\'a}n}, {Carpenter},
  {Wilner}, {Zhang}, {Zhu}, {Birnstiel}, {Bai}, {Benisty}, {Hughes},
  {{\"O}berg}, \& {Ricci}}]{2018ApJ...869L..41A}
{Andrews}, S.~M., {Huang}, J., {P{\'e}rez}, L.~M., {et~al.} 2018{\natexlab{a}},
  \apjl, 869, L41

\bibitem[{{Andrews} {et~al.}(2018{\natexlab{b}}){Andrews}, {Terrell},
  {Tripathi}, {Ansdell}, {Williams}, \& {Wilner}}]{2018ApJ...865..157A}
{Andrews}, S.~M., {Terrell}, M., {Tripathi}, A., {et~al.} 2018{\natexlab{b}},
  \apj, 865, 157

\bibitem[{{Andrews} {et~al.}(2012){Andrews}, {Wilner}, {Hughes}, {Qi},
  {Rosenfeld}, {{\"O}berg}, {Birnstiel}, {Espaillat}, {Cieza}, {Williams},
  {Lin}, \& {Ho}}]{2012ApJ...744..162A}
{Andrews}, S.~M., {Wilner}, D.~J., {Hughes}, A.~M., {et~al.} 2012, \apj, 744,
  162

\bibitem[{{Andrews} {et~al.}(2016){Andrews}, {Wilner}, {Zhu}, {Birnstiel},
  {Carpenter}, {P{\'e}rez}, {Bai}, {{\"O}berg}, {Hughes}, {Isella}, \&
  {Ricci}}]{2016ApJ...820L..40A}
{Andrews}, S.~M., {Wilner}, D.~J., {Zhu}, Z., {et~al.} 2016, \apjl, 820, L40

\bibitem[{{Ansdell} {et~al.}(2017){Ansdell}, {Williams}, {Manara}, {Miotello},
  {Facchini}, {van der Marel}, {Testi}, \& {van
  Dishoeck}}]{2017AJ....153..240A}
{Ansdell}, M., {Williams}, J.~P., {Manara}, C.~F., {et~al.} 2017, \aj, 153, 240

\bibitem[{{Ansdell} {et~al.}(2018){Ansdell}, {Williams}, {Trapman}, {van
  Terwisga}, {Facchini}, {Manara}, {van der Marel}, {Miotello}, {Tazzari},
  {Hogerheijde}, {Guidi}, {Testi}, \& {van Dishoeck}}]{2018ApJ...859...21A}
{Ansdell}, M., {Williams}, J.~P., {Trapman}, L., {et~al.} 2018, \apj, 859, 21

\bibitem[{{Ansdell} {et~al.}(2016){Ansdell}, {Williams}, {van der Marel},
  {Carpenter}, {Guidi}, {Hogerheijde}, {Mathews}, {Manara}, {Miotello},
  {Natta}, {Oliveira}, {Tazzari}, {Testi}, {van Dishoeck}, \& {van
  Terwisga}}]{2016ApJ...828...46A}
{Ansdell}, M., {Williams}, J.~P., {van der Marel}, N., {et~al.} 2016, \apj,
  828, 46

\bibitem[{{Bae} {et~al.}(2018){Bae}, {Pinilla}, \&
  {Birnstiel}}]{2018ApJ...864L..26B}
{Bae}, J., {Pinilla}, P., \& {Birnstiel}, T. 2018, \apjl, 864, L26

\bibitem[{{Barenfeld} {et~al.}(2016){Barenfeld}, {Carpenter}, {Ricci}, \&
  {Isella}}]{2016ApJ...827..142B}
{Barenfeld}, S.~A., {Carpenter}, J.~M., {Ricci}, L., \& {Isella}, A. 2016,
  \apj, 827, 142

\bibitem[{{Barenfeld} {et~al.}(2017){Barenfeld}, {Carpenter}, {Sargent},
  {Isella}, \& {Ricci}}]{2017ApJ...851...85B}
{Barenfeld}, S.~A., {Carpenter}, J.~M., {Sargent}, A.~I., {Isella}, A., \&
  {Ricci}, L. 2017, \apj, 851, 85

\bibitem[{{Birnstiel} \& {Andrews}(2014)}]{birnstiel_14}
{Birnstiel}, T. \& {Andrews}, S.~M. 2014, \apj, 780, 153

\bibitem[{{Birnstiel} {et~al.}(2012{\natexlab{a}}){Birnstiel}, {Andrews}, \&
  {Ercolano}}]{2012A&A...544A..79B}
{Birnstiel}, T., {Andrews}, S.~M., \& {Ercolano}, B. 2012{\natexlab{a}}, \aap,
  544, A79

\bibitem[{{Birnstiel} {et~al.}(2012{\natexlab{b}}){Birnstiel}, {Klahr}, \&
  {Ercolano}}]{2012A&A...539A.148B}
{Birnstiel}, T., {Klahr}, H., \& {Ercolano}, B. 2012{\natexlab{b}}, \aap, 539,
  A148

\bibitem[{{Boneberg} {et~al.}(2018){Boneberg}, {Facchini}, {Clarke}, {Ilee},
  {Booth}, \& {Bruderer}}]{2018MNRAS.477..325B}
{Boneberg}, D.~M., {Facchini}, S., {Clarke}, C.~J., {et~al.} 2018, \mnras, 477,
  325

\bibitem[{{Cazzoletti} {et~al.}(2019){Cazzoletti}, {Manara}, {Liu}, {van
  Dishoeck}, {Facchini}, {Alcal{\`a}}, {Ansdell}, {Testi}, {Williams},
  {Carrasco-Gonz{\'a}lez}, {Dong}, {Forbrich}, {Fukagawa}, {Galv{\'a}n-Madrid},
  {Hirano}, {Hogerheijde}, {Hasegawa}, {Muto}, {Pinilla}, {Takami}, {Tamura},
  {Tazzari}, \& {Wisniewski}}]{2019arXiv190402409C}
{Cazzoletti}, P., {Manara}, C.~F., {Liu}, H.~B., {et~al.} 2019, arXiv e-prints
  [\eprint[arXiv]{1904.02409}]

\bibitem[{{Cieza} {et~al.}(2019){Cieza}, {Ru{\'{\i}}z-Rodr{\'{\i}}guez},
  {Hales}, {Casassus}, {P{\'e}rez}, {Gonzalez-Ruilova}, {C{\'a}novas},
  {Williams}, {Zurlo}, {Ansdell}, {Avenhaus}, {Bayo}, {Bertrang},
  {Christiaens}, {Dent}, {Ferrero}, {Gamen}, {Olofsson}, {Orcajo}, {Pe{\~n}a
  Ram{\'{\i}}rez}, {Principe}, {Schreiber}, \& {van der
  Plas}}]{2019MNRAS.482..698C}
{Cieza}, L.~A., {Ru{\'{\i}}z-Rodr{\'{\i}}guez}, D., {Hales}, A., {et~al.} 2019,
  \mnras, 482, 698

\bibitem[{{Clarke} {et~al.}(2018){Clarke}, {Tazzari}, {Juhasz}, {Rosotti},
  {Booth}, {Facchini}, {Ilee}, {Johns-Krull}, {Kama}, {Meru}, \&
  {Prato}}]{2018ApJ...866L...6C}
{Clarke}, C.~J., {Tazzari}, M., {Juhasz}, A., {et~al.} 2018, \apjl, 866, L6

\bibitem[{{Dullemond} {et~al.}(2018){Dullemond}, {Birnstiel}, {Huang},
  {Kurtovic}, {Andrews}, {Guzm{\'a}n}, {P{\'e}rez}, {Isella}, {Zhu}, {Benisty},
  {Wilner}, {Bai}, {Carpenter}, {Zhang}, \& {Ricci}}]{2018ApJ...869L..46D}
{Dullemond}, C.~P., {Birnstiel}, T., {Huang}, J., {et~al.} 2018, \apjl, 869,
  L46

\bibitem[{{Dutrey} {et~al.}(1998){Dutrey}, {Guilloteau}, {Prato}, {Simon},
  {Duvert}, {Schuster}, \& {Menard}}]{1998A&A...338L..63D}
{Dutrey}, A., {Guilloteau}, S., {Prato}, L., {et~al.} 1998, \aap, 338, L63

\bibitem[{{Facchini} {et~al.}(2017){Facchini}, {Birnstiel}, {Bruderer}, \& {van
  Dishoeck}}]{2017A&A...605A..16F}
{Facchini}, S., {Birnstiel}, T., {Bruderer}, S., \& {van Dishoeck}, E.~F. 2017,
  \aap, 605, A16

\bibitem[{{Fedele} {et~al.}(2017){Fedele}, {Carney}, {Hogerheijde}, {Walsh},
  {Miotello}, {Klaassen}, {Bruderer}, {Henning}, \& {van
  Dishoeck}}]{2017A&A...600A..72F}
{Fedele}, D., {Carney}, M., {Hogerheijde}, M.~R., {et~al.} 2017, \aap, 600, A72

\bibitem[{{Foreman-Mackey} {et~al.}(2013){Foreman-Mackey}, {Hogg}, {Lang}, \&
  {Goodman}}]{2013PASP..125..306F}
{Foreman-Mackey}, D., {Hogg}, D.~W., {Lang}, D., \& {Goodman}, J. 2013, \pasp,
  125, 306

\bibitem[{{Furlan} {et~al.}(2011){Furlan}, {Luhman}, {Espaillat}, {D'Alessio},
  {Adame}, {Manoj}, {Kim}, {Watson}, {Forrest}, {McClure}, {Calvet}, {Sargent},
  {Green}, \& {Fischer}}]{2011ApJS..195....3F}
{Furlan}, E., {Luhman}, K.~L., {Espaillat}, C., {et~al.} 2011, \apjs, 195, 3

\bibitem[{{Gaia Collaboration} {et~al.}(2018){Gaia Collaboration}, {Brown},
  {Vallenari}, {Prusti}, {de Bruijne}, {Babusiaux}, {Bailer-Jones}, {Biermann},
  {Evans}, {Eyer}, {Jansen}, {Jordi}, \& et~al.}]{2018A&A...616A...1G}
{Gaia Collaboration}, {Brown}, A.~G.~A., {Vallenari}, A., {et~al.} 2018, \aap,
  616, A1

\bibitem[{{Gaia Collaboration} {et~al.}(2016){Gaia Collaboration}, {Prusti},
  {de Bruijne}, {Brown}, {Vallenari}, {Babusiaux}, {Bailer-Jones}, {Bastian},
  {Biermann}, {Evans}, {Eyer}, {Jansen}, \& et~al.}]{2016A&A...595A...1G}
{Gaia Collaboration}, {Prusti}, T., {de Bruijne}, J.~H.~J., {et~al.} 2016,
  \aap, 595, A1

\bibitem[{{Guilloteau} \& {Dutrey}(1998)}]{1998A&A...339..467G}
{Guilloteau}, S. \& {Dutrey}, A. 1998, \aap, 339, 467

\bibitem[{{Hartmann} {et~al.}(1998){Hartmann}, {Calvet}, {Gullbring}, \&
  {D'Alessio}}]{1998ApJ...495..385H}
{Hartmann}, L., {Calvet}, N., {Gullbring}, E., \& {D'Alessio}, P. 1998, \apj,
  495, 385

\bibitem[{{Hendler} {et~al.}(2017){Hendler}, {Mulders}, {Pascucci},
  {Greenwood}, {Kamp}, {Henning}, {M{\'e}nard}, {Dent}, \&
  {Evans}}]{2017ApJ...841..116H}
{Hendler}, N.~P., {Mulders}, G.~D., {Pascucci}, I., {et~al.} 2017, \apj, 841,
  116

\bibitem[{{Herczeg} \& {Hillenbrand}(2014)}]{2014ApJ...786...97H}
{Herczeg}, G.~J. \& {Hillenbrand}, L.~A. 2014, \apj, 786, 97

\bibitem[{{Huang} {et~al.}(2018){Huang}, {Andrews}, {Dullemond}, {Isella},
  {P{\'e}rez}, {Guzm{\'a}n}, {{\"O}berg}, {Zhu}, {Zhang}, {Bai}, {Benisty},
  {Birnstiel}, {Carpenter}, {Hughes}, {Ricci}, {Weaver}, \&
  {Wilner}}]{2018ApJ...869L..42H}
{Huang}, J., {Andrews}, S.~M., {Dullemond}, C.~P., {et~al.} 2018, \apjl, 869,
  L42

\bibitem[{{Hunter}(2007)}]{2007CSE.....9...90H}
{Hunter}, J.~D. 2007, Computing in Science and Engineering, 9, 90

\bibitem[{{Lodato} {et~al.}(2019){Lodato}, {Dipierro}, {Ragusa}, {Long},
  {Herczeg}, {Pascucci}, {Pinilla}, {Manara}, {Tazzari}, {Liu}, {Mulders},
  {Harsono}, {Boehler}, {M{\'e}nard}, {Johnstone}, {Salyk}, {van der Plas},
  {Cabrit}, {Edwards}, {Fischer}, {Hendler}, {Nisini}, {Rigliaco}, {Avenhaus},
  {Banzatti}, \& {Gully-Santiago}}]{2019MNRAS.486..453L}
{Lodato}, G., {Dipierro}, G., {Ragusa}, E., {et~al.} 2019, \mnras, 486, 453

\bibitem[{{Long} {et~al.}(2018){Long}, {Pinilla}, {Herczeg}, {Harsono},
  {Dipierro}, {Pascucci}, {Hendler}, {Tazzari}, {Ragusa}, {Salyk}, {Edwards},
  {Lodato}, {van de Plas}, {Johnstone}, {Liu}, {Boehler}, {Cabrit}, {Manara},
  {Menard}, {Mulders}, {Nisini}, {Fischer}, {Rigliaco}, {Banzatti}, {Avenhaus},
  \& {Gully-Santiago}}]{2018ApJ...869...17L}
{Long}, F., {Pinilla}, P., {Herczeg}, G.~J., {et~al.} 2018, \apj, 869, 17

\bibitem[{{McMullin} {et~al.}(2007){McMullin}, {Waters}, {Schiebel}, {Young},
  \& {Golap}}]{2007ASPC..376..127M}
{McMullin}, J.~P., {Waters}, B., {Schiebel}, D., {Young}, W., \& {Golap}, K.
  2007, in Astronomical Society of the Pacific Conference Series, Vol. 376,
  Astronomical Data Analysis Software and Systems XVI, ed. R.~A. {Shaw},
  F.~{Hill}, \& D.~J. {Bell}, 127

\bibitem[{{Najita} \& {Bergin}(2018)}]{2018ApJ...864..168N}
{Najita}, J.~R. \& {Bergin}, E.~A. 2018, \apj, 864, 168

\bibitem[{{Najita} {et~al.}(2007){Najita}, {Strom}, \&
  {Muzerolle}}]{2007MNRAS.378..369N}
{Najita}, J.~R., {Strom}, S.~E., \& {Muzerolle}, J. 2007, \mnras, 378, 369

\bibitem[{{Pani{\'c}} {et~al.}(2009){Pani{\'c}}, {Hogerheijde}, {Wilner}, \&
  {Qi}}]{2009A&A...501..269P}
{Pani{\'c}}, O., {Hogerheijde}, M.~R., {Wilner}, D., \& {Qi}, C. 2009, \aap,
  501, 269

\bibitem[{{Pascucci} {et~al.}(2016){Pascucci}, {Testi}, {Herczeg}, {Long},
  {Manara}, {Hendler}, {Mulders}, {Krijt}, {Ciesla}, {Henning}, {Mohanty},
  {Drabek-Maunder}, {Apai}, {Sz{\H u}cs}, {Sacco}, \&
  {Olofsson}}]{2016ApJ...831..125P}
{Pascucci}, I., {Testi}, L., {Herczeg}, G.~J., {et~al.} 2016, \apj, 831, 125

\bibitem[{{Pi{\'e}tu} {et~al.}(2014){Pi{\'e}tu}, {Guilloteau}, {Di Folco},
  {Dutrey}, \& {Boehler}}]{2014A&A...564A..95P}
{Pi{\'e}tu}, V., {Guilloteau}, S., {Di Folco}, E., {Dutrey}, A., \& {Boehler},
  Y. 2014, \aap, 564, A95

\bibitem[{{Pinilla} {et~al.}(2012){Pinilla}, {Birnstiel}, {Ricci}, {Dullemond},
  {Uribe}, {Testi}, \& {Natta}}]{2012A&A...538A.114P}
{Pinilla}, P., {Birnstiel}, T., {Ricci}, L., {et~al.} 2012, \aap, 538, A114

\bibitem[{{Ricci} {et~al.}(2010){Ricci}, {Testi}, {Natta}, {Neri}, {Cabrit}, \&
  {Herczeg}}]{2010A&A...512A..15R}
{Ricci}, L., {Testi}, L., {Natta}, A., {et~al.} 2010, \aap, 512, A15

\bibitem[{{Rosenfeld} {et~al.}(2013){Rosenfeld}, {Andrews}, {Wilner},
  {Kastner}, \& {McClure}}]{2013ApJ...775..136R}
{Rosenfeld}, K.~A., {Andrews}, S.~M., {Wilner}, D.~J., {Kastner}, J.~H., \&
  {McClure}, M.~K. 2013, \apj, 775, 136

\bibitem[{{Rosotti} {et~al.}(2019){Rosotti}, {Tazzari}, {Booth}, {Testi},
  {Lodato}, \& {Clarke}}]{2019MNRAS.tmp.1141R}
{Rosotti}, G.~P., {Tazzari}, M., {Booth}, R.~A., {et~al.} 2019, \mnras, 1141

\bibitem[{{Simon} {et~al.}(2017){Simon}, {Guilloteau}, {Di Folco}, {Dutrey},
  {Grosso}, {Pi{\'e}tu}, {Chapillon}, {Prato}, {Schaefer}, {Rice}, \&
  {Boehler}}]{2017ApJ...844..158S}
{Simon}, M., {Guilloteau}, S., {Di Folco}, E., {et~al.} 2017, \apj, 844, 158

\bibitem[{{Strom} {et~al.}(1989){Strom}, {Strom}, {Edwards}, {Cabrit}, \&
  {Skrutskie}}]{1989AJ.....97.1451S}
{Strom}, K.~M., {Strom}, S.~E., {Edwards}, S., {Cabrit}, S., \& {Skrutskie},
  M.~F. 1989, \aj, 97, 1451

\bibitem[{{Tazzari} {et~al.}(2018){Tazzari}, {Beaujean}, \&
  {Testi}}]{2018MNRAS.476.4527T}
{Tazzari}, M., {Beaujean}, F., \& {Testi}, L. 2018, \mnras, 476, 4527

\bibitem[{{Tazzari} {et~al.}(2017){Tazzari}, {Testi}, {Natta}, {Ansdell},
  {Carpenter}, {Guidi}, {Hogerheijde}, {Manara}, {Miotello}, {van der Marel},
  {van Dishoeck}, \& {Williams}}]{2017A&A...606A..88T}
{Tazzari}, M., {Testi}, L., {Natta}, A., {et~al.} 2017, \aap, 606, A88

\bibitem[{{Teague} {et~al.}(2018){Teague}, {Bae}, {Bergin}, {Birnstiel}, \&
  {Foreman-Mackey}}]{2018ApJ...860L..12T}
{Teague}, R., {Bae}, J., {Bergin}, E.~A., {Birnstiel}, T., \& {Foreman-Mackey},
  D. 2018, \apj, 860, L12

\bibitem[{{Testi} {et~al.}(2014){Testi}, {Birnstiel}, {Ricci}, {Andrews},
  {Blum}, {Carpenter}, {Dominik}, {Isella}, {Natta}, {Williams}, \&
  {Wilner}}]{2014prpl.conf..339T}
{Testi}, L., {Birnstiel}, T., {Ricci}, L., {et~al.} 2014, Protostars and
  Planets VI, 339

\bibitem[{{Trapman} {et~al.}(2019){Trapman}, {Facchini}, {Hogerheijde}, {van
  Dishoeck}, \& {Bruderer}}]{2019arXiv190306190T}
{Trapman}, L., {Facchini}, S., {Hogerheijde}, M.~R., {van Dishoeck}, E.~F., \&
  {Bruderer}, S. 2019, arXiv e-prints [\eprint[arXiv]{1903.06190}]

\bibitem[{{Tripathi} {et~al.}(2017){Tripathi}, {Andrews}, {Birnstiel}, \&
  {Wilner}}]{2017ApJ...845...44T}
{Tripathi}, A., {Andrews}, S.~M., {Birnstiel}, T., \& {Wilner}, D.~J. 2017,
  \apj, 845, 44

\bibitem[{{van Terwisga} {et~al.}(2018){van Terwisga}, {van Dishoeck},
  {Ansdell}, {van der Marel}, {Testi}, {Williams}, {Facchini}, {Tazzari},
  {Hogerheijde}, {Trapman}, {Manara}, {Miotello}, {Maud}, \&
  {Harsono}}]{2018A&A...616A..88V}
{van Terwisga}, S.~E., {van Dishoeck}, E.~F., {Ansdell}, M., {et~al.} 2018,
  \aap, 616, A88

\bibitem[{{Weidenschilling}(1977)}]{1977MNRAS.180...57W}
{Weidenschilling}, S.~J. 1977, \mnras, 180, 57

\bibitem[{{Zhang} {et~al.}(2018){Zhang}, {Zhu}, {Huang}, {Guzm{\'a}n},
  {Andrews}, {Birnstiel}, {Dullemond}, {Carpenter}, {Isella}, {P{\'e}rez},
  {Benisty}, {Wilner}, {Baruteau}, {Bai}, \& {Ricci}}]{2018ApJ...869L..47Z}
{Zhang}, S., {Zhu}, Z., {Huang}, J., {et~al.} 2018, \apjl, 869, L47

\end{thebibliography}

\begin{acknowledgements}
We thank the anonymous referee for the comments that helped improve this letter. We are grateful to S. Andrews, L. Testi, and J. Williams for fruitful discussions. This paper makes use of the following ALMA data: ADS/JAO.ALMA\#2016.1.00715.S. ALMA is a partnership of ESO (representing its member states), NSF (USA), and NINS (Japan), together with NRC (Canada),  NSC and ASIAA (Taiwan), and KASI (Republic of Korea), in cooperation with the Republic of Chile. The Joint ALMA Observatory is operated by ESO, AUI/NRAO, and NAOJ. SF and CFM acknowledge an ESO fellowship. MT has been supported by the DISCSIM project, grant agreement 341137 funded by the European Research Council under ERC-2013-ADG and by the UK Science and Technology research Council (STFC). GR acknowledges support from the Netherlands Organisation for Scientific Research (NWO, program number 016.Veni.192.233). Astrochemistry in Leiden is supported by the Netherlands Research School for Astronomy (NOVA). This project has received funding from the European Union’s Horizon 2020 research and innovation program under the Marie Skłodowska-Curie grant agreement No 823823. All the figures were generated with the {\tt python}-based package {\tt matplotlib} \citep{2007CSE.....9...90H}.
\end{acknowledgements}

\begin{appendix}

\section{Fitting and MCMC results}

\begin{table*}
\centering
\caption{Radii enclosing X\% of continuum intensity ($R_{\rm dust}$) and $^{12}$CO $J$=2-1 integrated intensity ($R_{\rm gas}$). }
\begin{tabular}{lcccc}
\hline
 & $50\%$ & $68\%$ & $90\%$ & $95\%$\\
\hline
\smallskip
$R_{\rm dust}\,[\arcsec]$ & $0.0777^{+0.0014}_{-0.0013}$ & $0.1094^{+0.0026}_{-0.0023}$ & $0.2299^{+0.0199}_{-0.0176}$ & $0.2939^{+0.0292}_{-0.0258}$\\
$R_{\rm gas}\,[\arcsec]$ & $0.44^{+0.02}_{-0.02}$ & $0.59^{+0.02}_{-0.02}$ & $0.90^{+0.07}_{-0.07}$ & $ 1.09^{+0.09}_{-0.09}$\\
$R_{\rm dust}\,$[AU] & $9.93^{+0.18}_{-0.17}$ & $13.99^{+0.33}_{-0.29}$ & $29.40^{+2.54}_{-2.25}$ & $37.59^{+3.73}_{-3.30}$\\
$R_{\rm gas}\,$[AU] & $56.3^{+2.6}_{-2.6}$ & $75.5^{+2.7}_{-2.7}$ & $115.1^{+8.9}_{-8.9}$ & $139.4^{+11.5}_{-11.5}$\\
\hline
\hline
\end{tabular}
\label{tab:radii}
\end{table*}

\begin{table*}
\centering
\caption{Median of the marginalized posteriors of the fitted parameters, with associated statistical uncertainties from  the 16th and 84th percentiles of the marginalized distributions. The center of the emission is computed with respect to the phase center of the visibilities, i.e., RA (ICRS) = 04:14:47:87352 and Dec (ICRS) = +26:48:10.6275.}
\begin{tabular}{ccccccccc}
\hline
$\log{I_0}$ & $\log{I_1}$ & $\gamma$ & $\sigma_0$ & $\sigma_1$ & $i$ & P.A. & $\Delta$RA & $\Delta$Dec\\
$[\log\,$Jy/steradian$]$ & $[\log\,$Jy/steradian] &  & [$\arcsec$] & [$\arcsec$] & [$^\circ$] & [$^\circ$] & [$\arcsec$] & [$\arcsec$]\\
\hline
\smallskip
$10.30^{+0.03}_{-0.03}$ & $8.81^{+0.14}_{-0.13}$ & $0.46^{+0.06}_{-0.05}$ & $0.067^{+0.003}_{-0.003}$ & $0.19^{+0.02}_{-0.02}$ & $55.1^{+1.0}_{-1.0}$ & $66.2^{+1.3}_{-1.4}$ & $0.0063^{+0.0006}_{-0.0006}$ & $-0.0187^{+0.0005}_{-0.0005}$\\
\hline
\hline
\end{tabular}
\label{tab:fit}
\end{table*}

\begin{figure}
\center
\includegraphics[width=0.9\columnwidth]{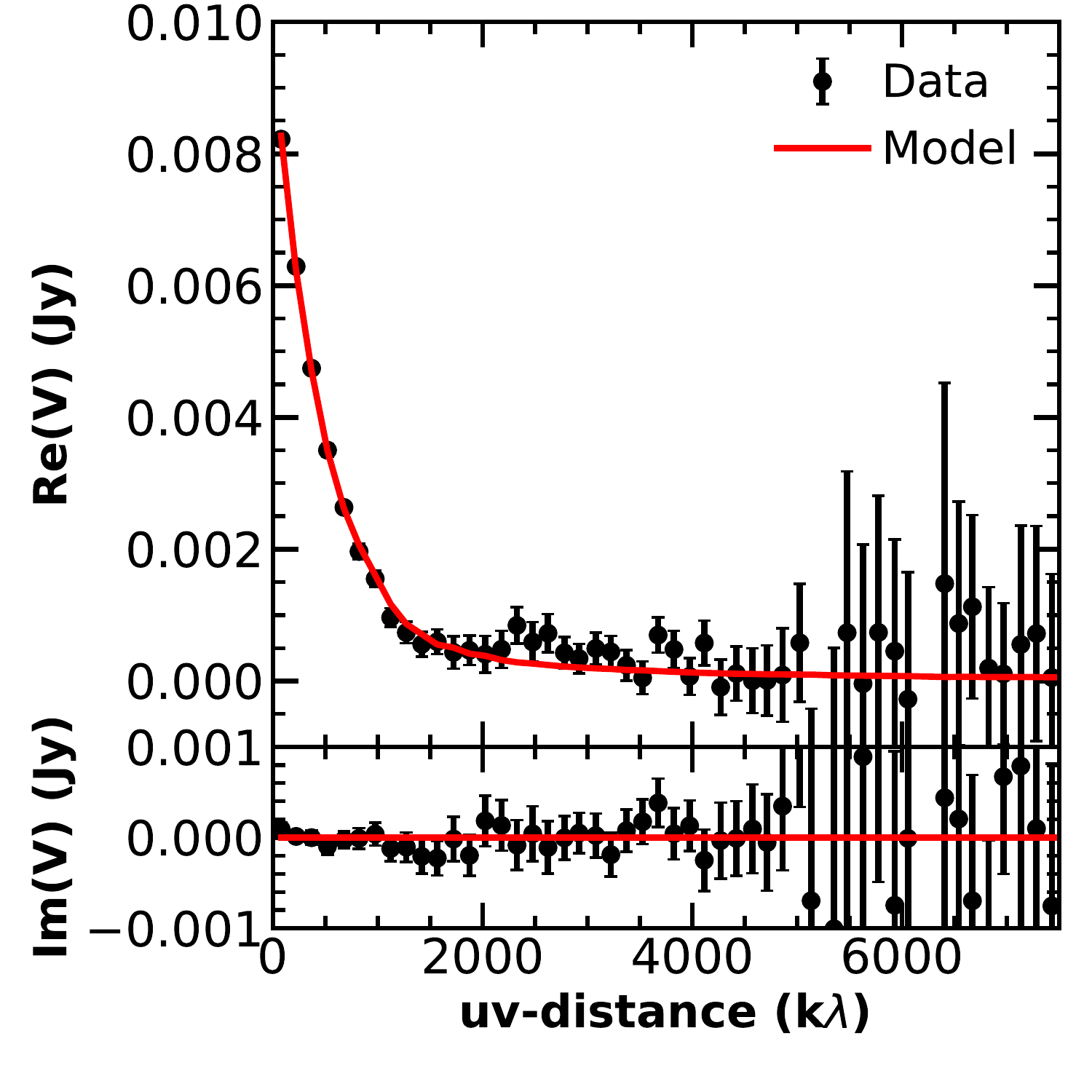}
\caption{Re-centered and de-projected visibilities and best fit model of the continuum data. Error bars are at $1\sigma$. The parameters of the best fit are listed in Table~\ref{tab:fit}.}
\label{fig:vis}
\end{figure}

\begin{figure*}
\begin{center}
\includegraphics[width=0.8\textwidth]{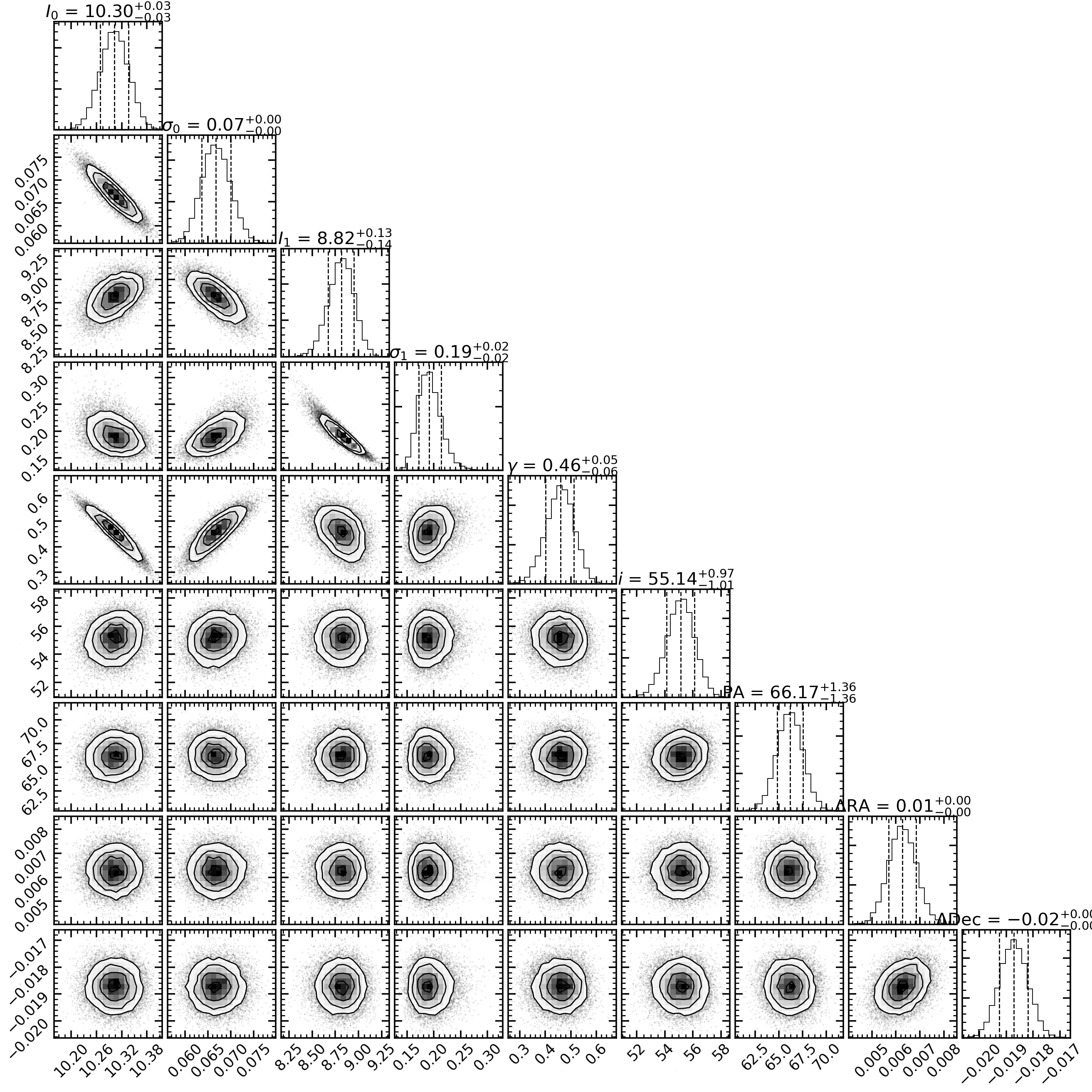}
\end{center}
\caption{Staircase plot of the chains of the MCMC fit without an inner cavity.}
\label{fig:triangle}
\end{figure*}

\begin{figure*}
\begin{center}
\includegraphics[width=0.8\textwidth]{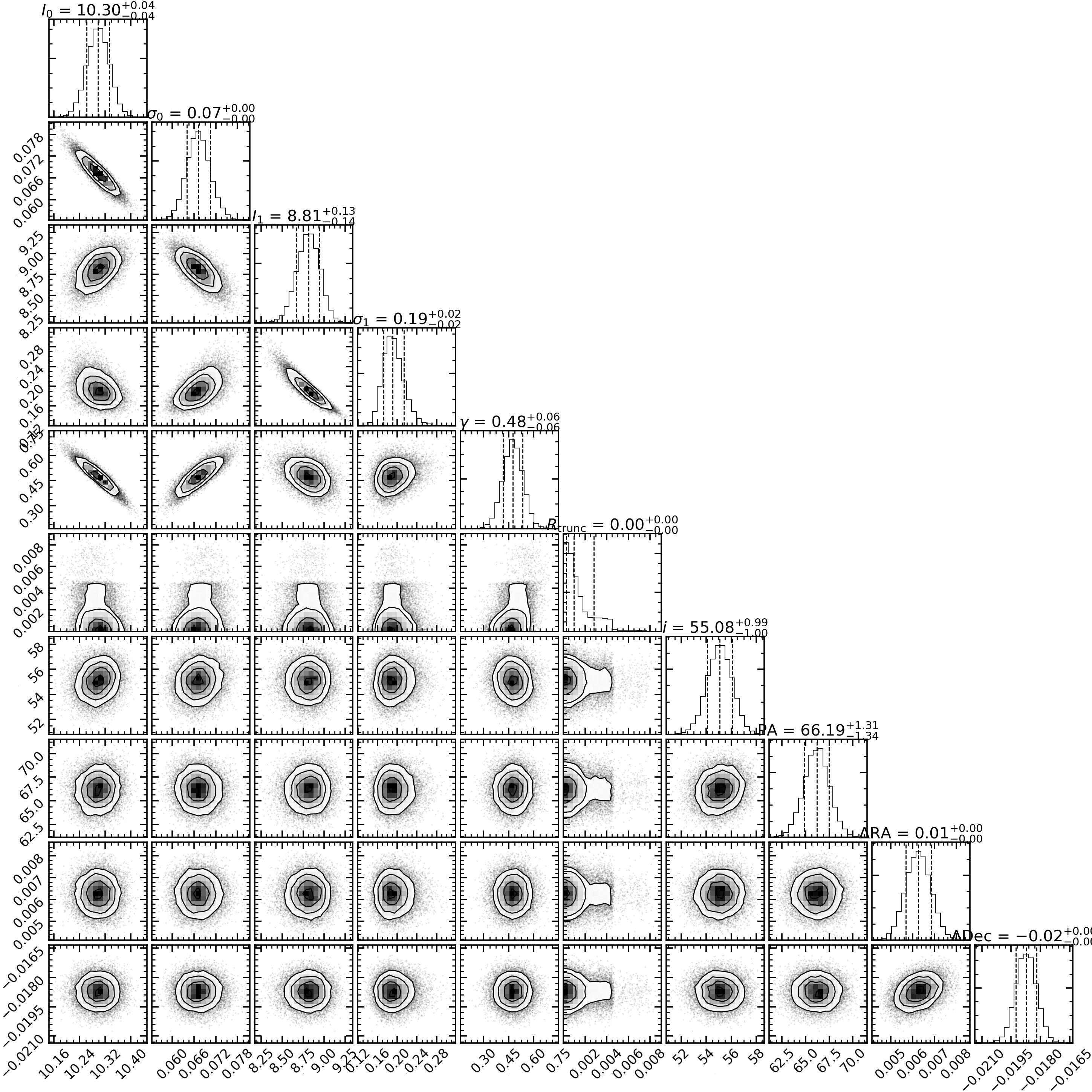}
\end{center}
\caption{Staircase plot of the chains of the MCMC fit with an inner cavity.}
\label{fig:triangle_2}
\end{figure*}

The fitting of the continuum intensity profile is performed using the GALARIO code \citep{2018MNRAS.476.4527T}. The code takes any model intensity profile (or 2D map), and computes the corresponding synthetic visibilities with the same $uv$-locations as in the ALMA data. Since the CX Tau disk is quite compact, and there are no evident azimuthal asymmetries, we model the flux density assuming azimuthal symmetry. A best fit model is then obtained by performing a fit using an MCMC sampler with the {\tt emcee} package \citep{2013PASP..125..306F}, with 90 walkers and 18000 steps after a burn in of 2000 steps. To fit the intensity profile with the addiction of a central cavity, 100 walkers are used (see Section~\ref{sec:res_3}).  We assume uniform priors for the fitted parameters, with the surface brightness normalization parameters are sampled logarithmically: $p(\log{I_0/({\rm Jy/steradian})})=\mathcal{U}[0,20]$; $p(\log{I_1/({\rm Jy/steradian})})=\mathcal{U}[0,20]$; $p(\gamma)=\mathcal{U}[0,3]$; $p(\sigma_0)=\mathcal{U}[0,0.2\arcsec]$; $p(\sigma_1)=\mathcal{U}[0.1,0.35\arcsec]$, with the restriction of $\sigma_1>\sigma_0$; $p(i)=\mathcal{U}[0,90^\circ]$; $p({\rm P.A.})=\mathcal{U}[0,180^\circ]$; $p(\Delta {\rm RA})=\mathcal{U}[-0.2,0.2\arcsec]$; $p(\Delta {\rm Dec})=\mathcal{U}[-0.2,0.2\arcsec]$; $p(R_{\rm trunc})=\mathcal{U}[0,0.1\arcsec]$, with the restriction of $R_{\rm trunc}<\sigma_0$. In both cases, the uncertainties are computed from independent posterior samples obtained by thinning the MCMC chains. The best fit model when no inner cavity is included is shown in Fig.~\ref{fig:vis}. The median of the marginalized distribution of the fitted parameters from Section~\ref{sec:results_cont} with relative statistical uncertainties are listed in Table \ref{tab:fit}. The corner plots of the two fits are shown in Figs.~\ref{fig:triangle} and \ref{fig:triangle_2}.

\end{appendix}

\end{document}